\documentstyle[preprint,aps,psfig]{revtex}
\begin{document}
\draft
\title{Solid-to-solid isostructural transition in the
hard sphere/attractive Yukawa system}

\author{C. Rasc\'{o}n$^{1,2}$,L. Mederos$^2$ and G. Navascu\'{e}s$^{1,2}$}
\address{$^1$Departamento de F\'{\i}sica de la Materia
Condensada,
Universidad Aut\'{o}noma, Cantoblanco, Madrid E-28049,Spain}
\address{$^2$Instituto de Ciencia de Materiales (Consejo
Superior de
Investigaciones Cient\'{\i}ficas), Cantoblanco, Madrid E-28049,
Spain}
\date{\today}
\maketitle

\begin{abstract}
A thermodynamically consistent density functional-perturbation
theory is used to study the isostructural solid-to-solid transition
which takes place in the hard sphere/attractive Yukawa system when
the Yukawa tail is sufficiently short-ranged. A comparison with results
for the square well potential allows us to study the effect of the
attractive potential form on the solid-solid transition. Reasonable
agreement with simulations is found for the main transition
properties as well as for the phase diagram evolution with the
the range of the attractive potential.
\end{abstract}

\pacs{64.70.Kb, 61.20.Ja, 82.70.Dd}

\section{introduction}

The phase diagrams of colloidal suspensions have become a subject of
growing interest in the last few years. Recent experimental
results for mixtures of spherical colloidal particles
and nonadsorbing polymers \cite{Leal,Pusey94} show very interesting
behaviour when the size of the polymer is reduced: The liquid-vapour
critical temperature decreases and, eventually, it becomes
smaller than the triple point temperature giving rise to a
phase diagram where only the fluid and solid phases are stable and
no liquid-vapour transition appears. A
theoretical understanding of this behaviour has been obtained from
simplified models which substitute the polymer by an effective
interaction which is attractive over distances larger than the
colloidal particle diameter $\sigma$. The origin of this attraction
is an entropic effect \cite{Meijer91}. However, the important point,
as far as the aim of this paper is concerned, is that
the range of this attraction is related to the polymer size and,
therefore, reducing the polymer size means reducing the range of
the attractive interaction between the colloidal particles.
This mechanism gives the possibility of producing real "simple" systems
with varying range of the interaction potential. Simple
expressions (like the square-well potential or hard spheres with an
attractive Yukawa tail) have been used for the resulting effective
potential between colloidal particles in theoretical
\cite{MederosHSAY,RasconSW} and computer simulation \cite{Meijer94,Bolhuis}
studies of the liquid phase stability as a function of the range of
the attractive interaction. Reasonable agreement between theory and
simulation have been found (see figures [1-3] in ref. \cite{MederosHSAY})
and the experimental results are qualitatively well understood. In
summary, it is concluded that when the range of the attractive tail
is smaller than $\approx \sigma/3$ the liquid phase is not present in
the phase diagram. In this case, when the gas density is increased at
fixed temperature, the system reduces its free energy transforming
itself into a solid before a van der Waals loop is generated.
The sublimation line is then always
above the liquid-gas coexistence curve in the density-temperature
phase diagram and the system only shows solid and fluid as stable
phases.

Even more interesting phenomena take place if the range $\delta$ of the
attractive tail is further reduced below $\sigma/3$. Although
direct experimental evidence does not exist so far, both simulations
\cite{Bolhuis} and theory \cite{RasconSW,Tejero,Likos1,Likos2} indicate that
when $\delta$ is small enough an isostructural solid-to-solid transition
takes place between an expanded solid and a condensed one. This novel
transition has some common features with the gas condensation, although
its mechanism is different \cite{RasconSW}. It is a first order transition
that takes place between two phases with the same structure. It ends
at a critical point as the temperature is increased and its
location (in the density-temperature plane, for example) strongly
depends on the range $\delta$ of the attractive tail.
However, some qualitative differences with respect to the condensation
are rapidly observed. For example, the solid-solid transition is strongly
asymmetric around its critical density and, contrary to the gas
condensation, the qualitative way in which its critical temperature
behaves with $\delta$ crucially depends on the {\it form} of the
attractive tail. In this way, $T_c$ remains practically constant as $\delta$
is varied for the square well (SW) system and decreases with $\delta$
for the hard sphere/attractive Yukawa potential (HSAY).

In this paper we report theoretical predictions for the solid-solid transition
of
the HSAY potential,

\begin{eqnarray}
\varphi (r) = \left \{ \begin{array}{ll}
\infty &  ,r < \sigma \\
- \epsilon {{exp[\kappa\sigma(1-r/\sigma)]} \over {r/\sigma}} &
,r > \sigma, \end{array} \right.
\label{potential}
\end{eqnarray}
obtained from our recently proposed density functional-perturbation
theory \cite{Mederos9394}. In this work we will focus on extremely
short-ranged attractive potentials ($\kappa\sigma \geq 25$). As already
pointed out, this theory gives reasonable
agreement with simulations for smaller $\kappa$ values ($\kappa\sigma
\leq 6$) where the phase diagram shows its usual aspect with
gas, liquid and solid stable phases. It is also able to predict reasonably
the transition to a phase diagram without stable liquid phase
for larger $\kappa\sigma$ values (see ref \cite{MederosHSAY}).  As
shown below, our results for the solid-solid transition are also in
qualitative agreement with simulations. Therefore, our density
functional-perturbation theory
gives a complete description of the phase diagram evolution with the
range of the attractive interaction.

The remainder of the paper is arranged as follows. In section II we
briefly summarize our theory. We present and discuss our results in
section III where we also include a comparison with simulations and
with the predictions from other theoretical models
\cite{Tejero,Likos1,Likos2}.

\section{DENSITY FUNCTIONAL-PERTURBATION THEORY for solids}

The density functional formalism has been successfully applied to
describe the solid phase of a hard sphere (HS) system in such
a way that we have today very accurate density functional approximations
for different properties of this system like, for example, its
melting transition or its structure when it is placed in front of
a hard wall, probably two of the strongest test for any theory
of non-uniform classical systems \cite{Evans}. All these
theories for non-uniform HS are based on some kind of mapping
into a uniform system at some effective density. Different recipes
to obtain the effective density from the physical one, $\rho({\bf r})$,
give rise to different density functional approximations. However, this
strategy will not work when it is applied to systems with more
realistic potentials, like (\ref{potential}), which include attractive
interactions between particles. The reason is that the thermodynamic
functions of the uniform fluid are not well defined inside the
density gap corresponding to condensation \cite{Kuijper,Kyrlidis91}.
It could be argued
\cite{Likos1,Likos2} that this is not an important problem when
we are interested in cases where, like the ones we study in this
paper, the liquid-gas transition is not present in the phase
diagram because the range of the attractive tail is very small.
However, there are reasons to believe that, even in these
cases, the mapping strategy is problematic \cite{RasconComent}. In
effect, we would like to have a theory able to describe the
complete phase diagram evolution with the potential range. It is
obvious that this is not possible with the mapping strategy. Moreover,
although the liquid-gas transition does not appear in the
phase diagram when the potential range is small because it is
preempted by the fluid-solid transition, the fluid free
energy branch with the usual van der Waals loop corresponding
to condensation does exist at low temperatures \cite{RasconSW}. Therefore,
the mapping strategy will fail in this temperature region.

An alternative is obtained from perturbation theories (PT) which
have been successfully used in uniform fluids \cite{Hansen86}.
The standard perturbation scheme provides a density functional
approximation for the Helmholtz free energy,
$F \left[\rho({\bf r})\right]$, of non-uniform systems. Its general
expression is:

\begin{equation}
F \left[\rho({\bf r})\right]= F_{ref} \left[\rho({\bf
r})\right] +
U_{p} \left[\rho({\bf r})\right].
\label{energy}
\end{equation}
$F_{ref}$ is the Helmholtz free energy of the reference system and
$U_p$ is the perturbation contribution to the free energy.
Up to first order, $U_p$ is given by

\begin{equation}
U_{p} \left[\rho({\bf r})\right]={1 \over 2} \int d{\bf r}
d{\bf r}^{'}
\rho^{(2)}_{ref}({\bf r},{\bf r}^{'})\varphi_p(| {\bf r}- {\bf
r}^{'}|),
\label{perturbation}
\end{equation}
where $\varphi_p(r)$ is the perturbation and
$\rho^{(2)}_{ref}({\bf r},{\bf r}^{'})$ is the pair distribution
function of the reference system. This function is usually written in
terms of $g_{ref}({\bf r},{\bf r}^{'})$,
the extension of the radial distribution function (rdf) to non-uniform
systems, as

\begin{equation}
\rho^{(2)}_{ref}({\bf r},{\bf r}^{'}) \equiv \rho({\bf r})
\rho({\bf r}^{'})
g_{ref}({\bf r},{\bf r}^{'}).
\label{ro2}
\end{equation}

The reference system is now mapped into an equivalent HS system of
diameter $d_{HS}$ whose free energy is obtained from any of the
accurate free energy functionals for non-uniform HS available in the
literature. We use Tarazona model \cite{Tarazona85} in this work
since it gives good results for the solid equation of state up to
very high densities ($\approx 1.4d_{HS}^3$). This is a required
condition to describe the solid-solid transition because the condensed
solid density reaches values near the close packing limit when the
temperature is low and/or the range of the attractive potential is
very small.  We do not need to specify a criterion to obtain the
equivalent HS diameter $d_{HS}$ from the reference potential in this
case because the natural way to split the potential
(\ref{potential}) into a reference and a perturbation parts
already gives a hard sphere system of diameter $\sigma$ as reference.
However, we need some approximation for the reference rdf
$g_{ref}({\bf r},{\bf r}^{'})$ because we know very little of it in
the non-uniform case. This problem is usually addressed making again
some mapping into the uniform limit at some effective density
$\hat \rho$:

\begin{equation}
g_{ref}({\bf r},{\bf r}^{'}) \approx g_u(|{\bf r}-{\bf r}^{'}|,\hat
\rho),
\label{gu}
\end{equation}
where $g_u(|{\bf r}-{\bf r}^{'}|,\rho)$ is the rdf of the uniform
reference (HS in this case) system at density $\rho$. The effective
density may in general depend on the position. It remains to give
some recipe to obtain $\hat \rho$. It is obvious that any
recipe based on the local density $\rho({\bf r})$ will not work in
the solid phase because $\rho({\bf r})$ reaches values which are
many times larger than the maximum allowed density in a uniform
fluid. A mapping like (\ref{gu}) has been used in PT for the
Lennard-Jones against a wall \cite{Tang,Soko} or for its melting
transition \cite{Kyrlidis93} using recipes for $\hat \rho$ which are the
{\it same} or similar than those used for the free energy
mapping. However, there are no reasons to think that the solid
correlation structure and its free energy can be mapped into the
uniform limit at equal or similar effective densities. It is
clear that the solid free energy mapping is obtained using effective
densities approximately equal to the {\it mean} solid density (this
is the result from current density functionals for HS). However,
the effective density we need in the rdf mapping (at least as far as the
{\it perturbation energy calculation} is concerned) is expected to be very
small (and then $g_u$ should be almost a step function). The reason
is that most of the correlation structure of $\rho^{(2)}$ in
(\ref{perturbation}) is {\it already} implicitly included in the
density product of equation (\ref{ro2}). To take into account this
idea we have recently proposed an alternative based on imposing an
{\it exact} sum rule to obtain $\hat \rho$ (see ref. \cite{Mederos9394}).
This is the {\it local compressibility} equation:

\begin{equation}
\int d{\bf r}^{'}\rho({\bf r}^{'})
[g_{ref}({\bf r},{\bf r}^{'})-1]=-1+
{k_BT\over{\rho({\bf r})}}{ d\rho({\bf r})\over{d\mu}},
\label{compre}
\end{equation}
where $\mu$ is the chemical potential. Precisely, $\hat{\rho}$
is defined
through equation (\ref{compre}) substituting the non-uniform rdf
$g_{ref}({\bf r},{\bf r}^{'})$ by its uniform limit
$g_{ref}(|{\bf r}- {\bf r}^{'}|,\hat{\rho}({\bf r}))$ and solving for
$\hat{\rho}({\bf r})$. A simpler version of this theory may be
obtained using a constant effective density $\hat \rho$ which is
obtained from the {\it exact} global compressibility equation
\cite{Hansen86}:

\begin{equation}
{{1} \over {N}}\int d{\bf r}\rho({\bf r})\int d{\bf r}^{'}\rho({\bf r}^{'})
[g_{ref}({\bf r},{\bf r}^{'})-1]=-1+k_BT\rho\chi_T,
\label{compreg}
\end{equation}
where $\chi_T$ is the solid isothermal compressibility and $\rho$ its mean
density. Using this recipe we obtain $\hat \rho \leq 0.08$. This is
consistent with our previous intuition on $g_u$ and also means that
the mean field approximation obtained using a step function for $g_u$
should be a reasonable approximation for the perturbation contribution to the
free energy of
the solid. Notice also that Eqn. (\ref{compre}) or (\ref{compreg})
reduces to the well known compressibility equation in the uniform
limit. Therefore, our theory reduces to the standard perturbation
theory in this limit and, in this way, we have a theory
which treats both phases on the same footing. This is crucial since
we are then able to determine consistently the relative stability of one
phase with respect to the others as it is required to study the
corresponding phase transitions and, therefore, to obtain the
phase diagram. Our theory is also able
to continuously go from one phase to the other if the local
compressibility equation (\ref{compre}) with a non-constant effective
density is used. This allows a consistent study of the solid-liquid
interface. Moreover, it does not suffer from the same problems than
the non-perturbative approximations (as discussed above) and, therefore,
it is able to describe the phase diagram for arbitrary values of the range
of the attractive potential. Then we can study the complete
phase diagram evolution using always the same theory. This is an
important point in order to give predictions for the relevant values of
$\kappa\sigma$  where the phase diagram suffers interesting qualitative
changes like, for example, the disappearance of the liquid phase or the
appearance of the solid-solid transition.

\section{Results and Discussion}

We have studied the phase diagram of the HSAY potential (\ref{potential})
for $\kappa\sigma =25$, $33$, $40$, $50$ and $67$ using the PT of the
previous section in its simplest form, i.e. using a constant effective
density $\hat \rho$ obtained from the global compressibility equation
(\ref{compreg}). The election of these particular values of
$\kappa\sigma$ has been motivated by the existence of simulation results
for those values \cite{Bolhuis}.
 We restrict ourselves to study the face-centered-cubic
(fcc) structure for the solid phase. As usual, we write the solid local
density, $\rho({\bf r})$,  as a sum over the lattice sites of normalized
gaussian peaks. The free energy functional (\ref{energy})is minimized with
respect to the gaussian width parameter. The standard common tangent method
is then used to obtain the coexisting densities of the different phases.
 The same theoretical approach was used in our previous
work \cite{MederosHSAY} on the same potential where we studied the
liquid phase extinction as $\kappa\sigma$ is increased and the attractive
potential becomes more and more short-ranged. Therefore, those results
together with the ones we are presenting in this section show a complete
and consistent picture of the phase diagram evolution.

Our results are shown in figures 1-5 as continuous lines. The theoretical
results for transitions between metastable states are shown by dashed
lines. We also show the simulation results of Bolhuis {\it et al.}
\cite{Bolhuis} (dotted lines). Note that the simulation results for the
solid-liquid transition for the cases $\kappa\sigma =40$, $50$, and $67$
have not been reported. The main characteristics of the
solid-solid transition summarized in the Introduction are qualitatively
reproduced by the theory. Contrary to the SW case
\cite{RasconSW,Bolhuis}, the critical point temperature $T_c$ depends
now on the range of the attractive tail. In fact, both simulations and
theory indicate that $T_c$ is almost a linear function of $\kappa$.
We can give arguments to understand this difference:
The attractive energy per particle (in $k_BT$ units) may be reasonably
approximated, when the range of the attractive potential is small enough,
by

\begin{equation}
{{F_{att}} \over {Nk_BT}} \approx {{1} \over {2k_BT}}n_n
\int d{\bf r} \rho_{{\bf R_1}}({\bf r}) g_u(r,\hat\rho)
\varphi_p(r),
\label{fatt}
\end{equation}
where $\rho_{{\bf R_1}}({\bf r})$ is the {\it normalized} gaussian
density peak centred at the nearest neighbour shell distance ${\bf
R_1}$ and $n_n$ is the number of nearest neighbours (12 in the
fcc lattice). A further simplification can be introduced
considering the low values of $\hat\rho$ given by (\ref{compreg})
(as discussed in the previous section) and,
therefore, using the mean field approximation. Then $g_u$ is
approximated by a step function and Eqn (\ref{fatt}) can be
written as
\begin{equation}
{{F_{att}} \over {Nk_BT}} \approx {{6} \over {k_BT}}
\int d{\bf r} \rho_{{\bf R_1}}({\bf r})\varphi_p(r).
\label{fatt1}
\end{equation}
{}From this last equation it is easy to see that the form of $\varphi_p(r)$
plays an important role. For the SW at low solid densities
(i.e. those corresponding to the expanded solid) the gaussian peak
$\rho_{{\bf R_1}}({\bf r})$ is, roughly speaking, practically outside
the attractive well. Then the attractive energy is very small.
On the contrary, in the condensed solid side the gaussian peak
is completely inside the well. Then, since
the potential is constant and the gaussian peak is normalized, the
attractive energy is $\approx -6\epsilon / k_BT$. The important point
is that this result depends on the temperature only. It does {\it not}
depend on the well width nor on the density. Therefore, the critical
point temperature is not expected to depend on this two parameters.
However, the same argument does not apply to the HSAY potential. The
main reason is that the form of the potential does not allow to say
when the density peak is inside the well and when it is outside. The
angular integration in (\ref{fatt1}) can be straightforwardly done to
obtain the attractive energy in terms of a single radial integral
whose result will obviously depend on $\kappa$. At a given density
its absolute value decreases when $\kappa$ increases. The solid-
solid critical temperature should then be lower when the range of
the attractive potential is reduced. It is interesting to stress
the dependence of the critical temperature behaviour with the
well width on the potential form. This is qualitatively different
from the liquid-vapour critical temperature case. This last
temperature is {\it always} reduced when the well width is
reduced, irrespective of the potential form. The reason resides in
the different transition mechanisms. The solid-solid transition takes
place when a particle is able to see the
attractive wells due to its neighbours. The precise moment when
this happens depends on the well form. A liquid particle, however,
{\it always} sees the attractive wells of the surrounding particles.
Its attractive energy is given (in the mean field approximation)
by the total integral of the potential
and this always depends  on the potential width and form.

We turn now to a more detailed comparison with the simulation results.
The theoretically predicted solid-solid critical temperature shows
a practically linear dependence with $\kappa$ with a slope in
agreement with simulations. However, our result for $T_c$  is
systematically above the simulations with an overestimation around
$13.5\%$. Some authors have attributed this overestimation to the
mean field nature of the theory, in the same way that mean field
approximations overestimate the liquid-vapour critical temperature
\cite{Likos1}. On the contrary, we believe that the origin of this
discrepancy between theory and simulation is different. As discussed
in the previous section, the mean field approximation for the solid
phase takes into account most of the solid correlation
structure. Moreover, we have obtained an overestimation around
$59\%$ in our prediction for the solid-solid critical temperature
of the SW potential \cite{RasconSW} using exactly the
same theory. It is difficult to believe that correlations together
with the different form of the potential may be responsible for this
enormous difference. Instead, we think that the solid-solid
transition is the result of the very delicate balance between the
different contributions to the free energy of the solid phase
\cite{RasconSW}. In fact, the final result crucially depends on particular
details of the model. A further confirmation is the fact that
we obtain $k_BT_c /\epsilon \approx 2.7$ for the SW potential
using the mean field approximation (i. e. using a step function
for $g_{ref}$) for the attractive energy together
with Tarazona model for the HS reference system. This result should
be compared with $k_BT_c /\epsilon \approx 2.0$ obtained by
Likos {\it et al.} using the very same mean field approximation for the
attractive energy but using Denton and Ashcroft functional for HS
\cite{Denton}. Moreover, Likos {\it et al.} have studied the
solid-solid transition of the SW potential using a
non-perturbative approximation for the full potential \cite{Likos2}.
While they get improved results for the critical temperature as
compared with their previous mean field calculation \cite{Likos1},
they obtain now a critical temperature which depends on the potential
range, in opposition to simulation and mean field results.

Fig. 1-5 show that our theoretical result for the critical solid-
solid density is slightly smaller than the simulation result.
However, this underestimation reduces when $\kappa$ increases. We
also overestimates the value of $\kappa$ where the solid-solid
transition emerges in the phase diagram (see figures 1-3).  It is
interesting to note that the predictions from the same theory for
these two properties in the
SW case are practically identical to those of simulations
\cite{RasconSW}. As already discussed,
the form of the potential makes now the transition mechanism less
decisive in order to fix the precise localization of the transition.
There is not an absolute criterion to say when a gaussian density
peak is inside the attractive wells due to its nearest neighbours and
when it is outside. This means that, contrary to the SW case,
the results for the relevant densities (like the critical and
triple points densities) will now depend on the model details because
the nearest neighbour distance is only function of the mean solid density.
However, this problem becomes less important when $\kappa$
increases because the solid-solid transition moves to higher densities,
near the close packing limit, where everything is geometrically fixed.

In conclusion, we find a solid-solid transition whose coexistence curve is,
as far as its general form is concerned, in good agreement with
simulation (see figures 3-5) but shifted in the density-temperature
plane. The results for the relevant properties (critical temperature and
density, etc.) depend on model details because the solid-solid
transition is the consequence of a delicate balance between the hard core
energy and the attractive energy. At the same time, this means that the
form of the attractive potential plays an important role which
determines some qualitative aspects like the dependence of the
critical temperature with the range of the potential. We have shown that
our perturbation theory, outlined in section II, is able to describe reasonably
the phase diagram of the HSAY system for the range of the
attractive potential ranging from "usual" values (where the phase diagram
shows liquid-gas, liquid-solid and vapour-solid transitions) to extremely
short-ranged potentials (where the liquid-vapour condensation is not
present) and even shorter-ranged potentials where the novel isostructural
solid-solid transition emerges in the phase diagram. To our knowledge,
this is the only consistent theory able to perform it successfully.

\acknowledgments

We thank Dr. C.N. Likos for sending us a preprint
of ref. \cite{Likos2} prior to publication.
This work was supported by the Direcci\'{o}n General de Investigaci\'{o}n
Cient\'{\i}fica
y T\'{e}cnica of Spain, under Grant $PB91-0090$.

\begin{figure}
\caption{Phase diagram of the hard sphere/attractive Yukawa potential in the
temperature-density plane for $\kappa\sigma = 25$. The solid lines are
the predictions of our theory. The dashed lines are also predictions
from our theory but for transitions between metastable states.
The dotted lines are the simulation results of Bolhuis {\it et} al.}
\end{figure}

\begin{figure}
\caption{As figure 1 for $\kappa\sigma = 33$.}
\end{figure}

\begin{figure}
\caption{As figure 1 for $\kappa\sigma = 40$.}
\end{figure}

\begin{figure}
\caption{As figure 1 for $\kappa\sigma = 50$.}
\end{figure}

\begin{figure}
\caption{As figure 1 for $\kappa\sigma = 67$.}
\end{figure}

\end{document}